\documentclass[aps,prl,floatfix,twocolumn,showpacs]{revtex4}
\usepackage{epsfig}

\usepackage{amssymb}
\usepackage{bbold}
\usepackage{amsmath}

\usepackage{color}
\usepackage{ulem}

\newcommand{\Tc}[3]{{\cal T}_{\,#2#1}^{#3}}

\begin{document}


\hsize\textwidth\columnwidth\hsize\csname@twocolumnfalse\endcsname

\title{Measuring Spin Accumulations with Current Noise}

\author{Jonathan Meair$^1$, Peter Stano$^{1,2}$ and Philippe Jacquod$^{1,3}$
\\ \textit{$^1$Physics Department, University of Arizona, 
1118 East Fourth Street, Tucson, AZ 85721, USA \\
$^2$Institute of Physics, Slovak Academy of Sciences, Bratislava 845 11, Slovakia\\
$^3$College of Optical Sciences, University of Arizona, 
1630 East University Boulevard, 
Tucson, AZ 85721, USA}}

\vskip1.5truecm
\begin{abstract}
We investigate the time-dependent fluctuations of the electric current 
injected from a reservoir with a non-equilibrium spin accumulation
into a mesoscopic conductor. We show how the current noise power
directly reflects the magnitude of the spin accumulation 
in two easily noticeable ways.
First, as the temperature is lowered, the small-bias noise saturates at a 
value determined by the spin accumulation. Second, 
in the presence of spin-orbit interactions in the conductor, the current noise
exhibits a sample-dependent mesoscopic 
asymmetry under reversal of the electric current direction. 
These features provide for a purely 
electric protocol for measuring spin accumulations. 
\end{abstract}
\maketitle

Noise measurements on non-equilibrium electric currents are very efficient 
probes of the 
dynamics and nature of the charge carriers~\cite{bla00}. At low temperature,
the classical Johnson-Nyquist noise is suppressed and quantum effects
govern the behavior of the surviving shot noise. In the mesoscopic
regime, the noise
power $S$ 
is reduced below its uncorrelated Poisson value $S_0 = 2 |q| 
\langle I \rangle$,
where $\langle I \rangle$ is the average electric current, by the
Fano factor $F=S/S_0 $. The value of $F$
depends on the electronic dynamics. For instance,
one finds $F=1/3$ in diffusive systems and $F=1/4$ in ballistic
chaotic systems~\cite{bla00,bee03}.
Alternatively, shot noise measurements have determined the charge $|q|$ 
of current-carrying quasiparticles in normal-metal/superconductor junctions
and in the fractional quantum Hall effect~\cite{bla00,bee03,rez98}. In this
manuscript we further illustrate the usefulness of current noise measurements
by showing how they can reveal the magnitude of non-equilibrium spin 
accumulations. Our results provide for a purely electric protocol
to measure spin accumulations, which has the potential to 
quantitatively determine their {\it magnitude}. It therefore
goes one step further than the optical 
methods used so far to detect 
magneto-electrically generated spin
accumulations~\cite{she1,she2}. Alternatively, the noise measurement
we propose, coupled with an electric measurement of the
spin Hall and inverse spin Hall effects~\cite{she3,she4,she5}, 
can provide key experimental
information on the conversion between spin accumulations and spin currents.

\begin{figure}[h]
\includegraphics[width=8cm]{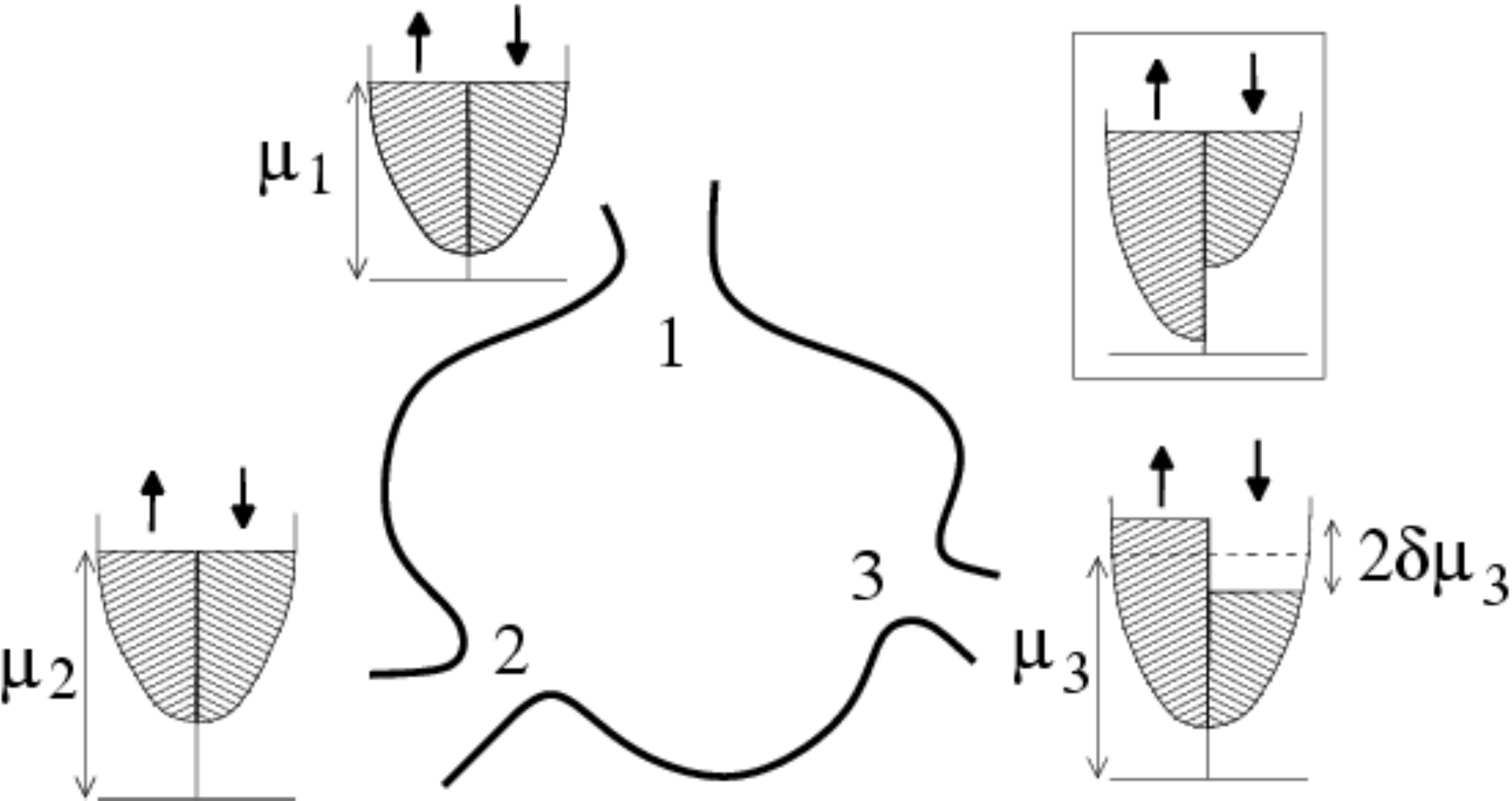}
\caption{\label{fig1} Three-terminal quantum dot connected to two unpolarized 
electron reservoirs (labeled 1 and 2) 
and one reservoir (3) with a non-equilibrium spin accumulation. Top right inset: a ferromagnetic
reservoir with an equilibrium spin accumulation (a case we do not consider in this paper).\\[-9mm]}
\end{figure}

A number of works have investigated charge current noise
from polarized reservoirs.
Reference \cite{erl05} suggested using current and 
noise measurements in the single-channel limit to measure 
the spin injection efficiency from a ferromagnet for weak 
spin flip scattering. 
Other related works have pointed out that noise measurements in
hybrid paramagnetic/ferromagnetic structures can reveal 
information on the 
relative orientation of the ferromagnets~\cite{tser} and
on the spin relaxation processes in the paramagnet~\cite{Mish,lam,belzig,Nag}. 
These results have been at least partially confirmed by numerical
simulations~\cite{nik}. In non-interacting systems, current cross-correlations
have a sign determined by the statistics of the charge carriers. 
Investigations of a single-level interacting fermionic 
quantum dot coupled to ferromagnetic leads 
have demonstrated the
emergence of positive (boson like) current
cross-correlations for certain relative orientations of 
the polarizations~\cite{bruder}. 
In all these instances, only 
ferromagnetic, i.e., ~equilibrium polarizations were considered. 
Below we show that non-equilibrium
spin accumulations
generate fundamentally different electric current noises.
Our main findings are that (i) at low enough temperature,
the small-bias noise saturates at a 
value reflecting the spin accumulation, and (ii) 
in the presence of spin-orbit interactions, the current noise
exhibits a sample-dependent, mesoscopic 
asymmetry under reversal of the electric current direction. These
two features appear only
in the presence of non-equilibrium spin accumulations. 

We consider a system such as the one 
sketched in Fig.~\ref{fig1}, where a mesoscopic
conductor is connected via multichannel leads to $M$ external reservoirs, 
$\alpha=1,2,\ldots, M$, at 
electro-chemical potentials $\mu_\alpha = (\mu_{\alpha \uparrow}+
\mu_{\alpha \downarrow})/2$ and with
non-equilibrium spin accumulations 
$\delta \mu_\alpha =(\mu_{\alpha \uparrow}-\mu_{\alpha \downarrow})/2$, 
along reservoir-dependent axes de- 
\begin{widetext}
\noindent
fined by unit vectors ${\bf m}_\alpha=(m_{\alpha x},m_{\alpha y},m_{\alpha z})$.
We use the linear response scattering approach to transport to write 
the zero-frequency noise power in units of $(e^2/h)$ as~\cite{bla00,nik} 
\begin{eqnarray}\label{eq:noise}
{S}_{\alpha \beta} &=&
\sum_{\gamma \delta} \sum_{(m,\sigma) \in \gamma}
\sum_{(n,\sigma') \in \delta} \int dE \;
A_{\gamma \delta}^{m,\sigma ;n,\sigma'} \left(\alpha;E\right) A_{\delta \gamma}^{n,\sigma';m,\sigma}\left(\beta;E\right) \;
\left[ 
f_\gamma^{\sigma}\left(1-f_\delta^{\sigma'}\right)+f_\delta^{\sigma'}\left(1-f_\gamma^{\sigma}\right) 
\right] \, ,
\end{eqnarray}
where $f_\gamma^{\sigma}$ is the Fermi
function for electrons with spin $\sigma=\pm$ along ${\bf m}_\gamma$ in
terminal $\gamma$, and
the sums run over all terminals $\gamma$ and $\delta$ (including
$\alpha$ and $\beta$), all channels $m \in \gamma$ and
$n \in \delta$, and all spin orientations 
$\sigma,\sigma'= \pm$. We defined 
\begin{eqnarray}
A_{\gamma \delta}^{m,\sigma ;n,\sigma'} \left(\alpha;E\right)
&= 
\delta_{mn} \delta_{\sigma \sigma'} \delta_{\alpha \gamma}
\delta_{\alpha \delta} 
-[s_{\alpha\gamma}^\dagger(E) s_{\alpha\delta}(E)]_{m,\sigma ;n,\sigma'} \, ,
\end{eqnarray}
where $s_{\alpha\gamma}$ denotes the $2 N_\alpha \times 2 N_\gamma$ subblock of
the scattering matrix of the total system, corresponding
to scattering from lead $\gamma$ to lead $\alpha$, $N_{\alpha,\gamma}$ 
being the number of channels in those leads. This assumes that
$N_{\alpha}$ is spin-independent in all leads, and we will comment on the
case $N_{\alpha \uparrow} \ne N_{\alpha \downarrow}$ later.
Equation (\ref{eq:noise}) differs from Eq.~(52) in Ref.~\cite{bla00} in that 
spin indices are explicitly written down here.
All our calculations below 
are current-conserving, gauge invariant, and satisfy linear
response reciprocity relations, as they should.
\end{widetext}
We assume that the temperature, applied voltages, and spin accumulations are low enough that the scattering matrix
is essentially constant in the energy interval where the 
square bracket in Eq.~(\ref{eq:noise}) does not vanish. We then 
substitute $A_{\gamma \delta}^{m,\sigma ;n,\sigma'} \left(\alpha;E\right)
\rightarrow 
A_{\gamma \delta}^{m,\sigma ;n,\sigma'} \left(\alpha;E_{\rm F}\right)$,
define
\begin{eqnarray}\label{eq:F}
{\cal F}^{\sigma \sigma'}_{\gamma \delta} 
\equiv 
\int dE  
\left[
f_\gamma^{\sigma}\left(1-f_\delta^{\sigma'}\right)+f_\delta^{\sigma'}\left(1-f_\gamma^{\sigma}\right) 
\right] \; , 
\end{eqnarray}
and introduce the two-terminal symmetry coefficients 
\begin{subequations}\label{eq:FAS}
\begin{eqnarray}
{\cal F}^{S S}_{\gamma \delta} 
& = &
\frac{1}{4}\sum_{\sigma \sigma'}
{\cal F}^{\sigma \sigma'}_{\gamma \delta} \; , \;\;
{\cal F}^{A A}_{\gamma \delta} 
=
\frac{1}{4}\sum_{\sigma \sigma'}
\sigma \sigma'
{\cal F}^{\sigma \sigma'}_{\gamma \delta}\; , \qquad  \\
{\cal F}^{A S}_{\gamma \delta} 
&=&
\frac{1}{4}\sum_{\sigma \sigma'}
\sigma
{\cal F}^{\sigma \sigma'}_{\gamma \delta} 
 \; , \;\;
{\cal F}^{S A}_{\gamma \delta} 
= 
\frac{1}{4}\sum_{\sigma \sigma'}
\sigma'
{\cal F}^{\sigma \sigma'}_{\gamma \delta}  \; . \qquad 
\end{eqnarray}
\end{subequations}
The indices $S$ ($A$) indicate that the function is symmetric
(antisymmetric) with respect to the spin accumulation in the
corresponding lead, e.g.~
${\cal F}^{S A}_{\gamma \delta}(\delta \mu_\gamma, 
\delta \mu_\delta) = {\cal F}^{S A}_{\gamma \delta}(-\delta \mu_\gamma, 
\delta \mu_\delta)=-{\cal F}^{S A}_{\gamma \delta}(\delta \mu_\gamma, 
-\delta \mu_\delta)$. We obtain
\begin{eqnarray}\label{eq:mother}
{S}_{\alpha \beta}
&=& 
2k_{\rm B} T \,
\left[2N_\alpha\delta_{\alpha\beta} - \text{Tr}\left(s_{\beta\alpha}^\dagger s_{\beta\alpha}+s_{\alpha\beta}^\dagger s_{\alpha\beta}\right)\right] \qquad 
\\
&+&
\sum_{\gamma \delta}
{\cal F}^{SS}_{\gamma \delta} \, 
\Tc{\alpha \beta}{\gamma \delta}{00}
+2 {\cal F}^{AS}_{\gamma \delta}
{\rm Re}
\Tc{\alpha \beta}{\gamma \delta}{z0} + 
{\cal F}^{AA}_{\gamma \delta} 
\Tc{\alpha \beta}{\gamma \delta}{zz}
\; , \nonumber 
\end{eqnarray}
with the spin-dependent noise coefficients 
\begin{eqnarray}
\label{eq:calS}
\Tc{\alpha \beta}{\gamma \delta}{ab}&=&
\text{Tr}\left[
\left(\mathbb{1}_\gamma\otimes\sigma_\gamma^a\right)
s_{\alpha\gamma}^\dagger s_{\alpha\delta}
\left(\mathbb{1}_\delta\otimes\sigma_\delta^b\right)
s_{\beta\delta}^\dagger s_{\beta\gamma}\right].
\end{eqnarray}
Here, the
trace runs over both spin and channel indices, $\mathbb{1}_\gamma$ is
the $N_\gamma \times N_\gamma$ identity matrix, 
$\sigma_\gamma^z \equiv \boldsymbol{\sigma} \cdot {\bf m}_\gamma$, 
where $\boldsymbol{\sigma}$ is the 
vector of Pauli matrices, and $\sigma_\gamma^0$ is
the $2 \times 2$ identity matrix. 
The coefficients given by Eq.~\eqref{eq:calS} generalize those introduced in Ref.~\cite{bar07} for the calculation of
spin conductance, to the calculation of noise. The linear
response Eq.~(\ref{eq:mother}) is valid for
any number of terminals whose temperatures, 
electro-chemical potentials, and spin accumulations
are encoded in the coefficients ${\cal F}$, and for any 
particle dynamics contained in the noise coefficients ${\cal T}$. 

We first mention symmetry properties of the coefficients ${\cal F}$. Aside from
their symmetry with respect to spin accumulations [see Eqs.~(\ref{eq:FAS})], 
they satisfy (i) ${\cal F}^{S A}_{\gamma \delta}={\cal F}^{A A}_{\gamma \delta}={\cal F}^{A S}_{\delta \gamma}=0$
if $\delta \mu_\delta=0$, (ii) 
${\cal F}^{S A}_{\gamma \delta}={\cal F}^{A S}_{\delta \gamma}$,
(iii) ${\cal F}^{S S}_{\gamma \delta}$ and ${\cal F}^{A A}_{\gamma \delta}$ 
are symmetric, while ${\cal F}^{S A}_{\gamma \delta}$ and 
${\cal F}^{A S}_{\gamma \delta}$ are antisymmetric 
with respect to the voltage bias between $\gamma$ and $\delta$, and
(iv) ${\cal F}^{A S}_{\gamma \gamma}=0$. 
Property (iii) is of particular interest, since together 
with Eq.~(\ref{eq:mother}), it implies 
that in the presence of spin-orbit interactions, {\it the noise power is no longer symmetric under reversal of the 
current/voltage when there is spin accumulation} in at least one reservoir. 

The system-dependent noise coefficients ${\cal T}$ are determined by the 
orbital and spin dynamics of the electrons. 
We calculate their mesoscopic ensemble average and, when it vanishes,
their typical value, taken as the root mean square of their distribution.
In the absence of spin accumulation,
only spin-independent coefficients
$\Tc{\alpha \beta}{\gamma \delta}{00}$ 
enter Eq.~\eqref{eq:mother}, whose
mesoscopic averages
$\langle \Tc{\alpha \beta}{\gamma \delta}{00} \rangle$ 
have been computed using, for example, random matrix theory~\cite{Bro96} or the trajectory-based semiclassical
theory~\cite{whi06}. Extended to account for the 
Pauli matrices in Eq.~(\ref{eq:calS}), these methods give for chaotic ballistic systems
\begin{equation}
\langle 
\Tc{\alpha \beta}{\gamma \delta}{ab} 
\rangle = 
2\frac{N_\alpha N_\beta N_\gamma N_\delta}{N_{\rm T}^2} \delta_{ab} \left[ \left(\frac{\delta_{\alpha\beta}}{N_\alpha}-\frac{1}{N_{\rm T}}\right)\delta_{a0} + \frac{\delta_{\gamma\delta}}{N_\delta}  \right],
\label{eq:chaotic}
\end{equation}
a result which holds to leading order in the total number of channels, 
$N_{\rm T}=\sum_\alpha N_\alpha \gg 1$, and
for both the unitary (broken time reversal symmetry) and the symplectic (broken spin rotational
symmetry but preserved time reversal symmetry) ensembles~\cite{mehta}. In the orthogonal ensemble (preserved spin rotational and time reversal symmetries), Eq.~\eqref{eq:chaotic} holds provided one substitutes $\delta_{a0} \rightarrow 1$. Then, in the case of non-collinear spin accumulations in leads $\gamma$ and  $\delta$, the symbol $\delta_{ab}$ for $a=z=b$ should be understood as ${\bf m}_\gamma \cdot {\bf m}_\delta$. 

As a first example, we consider a spin preserving system with only collinear spin accumulations. This gives 
$\Tc{\alpha \beta}{\gamma \delta}{zz}=\Tc{\alpha \beta}{\gamma \delta}{00}$ 
and 
$\Tc{\alpha \beta}{\gamma \delta}{z0} =0$. 
Only spin-diagonal coefficients ${\cal F}^{\sigma\sigma}$ 
enter Eq.~\eqref{eq:mother} and the two spin species are uncorrelated, 
with additive contributions to the current 
noise.
Despite zero charge current, the current noise can be finite in the presence of
spin accumulations.

Aiming at an all-electrical measurement protocol for spin accumulations, 
we show how the previous {\it a priori} trivial observation carries over to spin systems  
with fully broken spin rotational symmetry, where the electron dwell time 
is larger than the spin-orbit time. 
For simplicity, we focus on symmetric two-terminal geometries, $N=N_L=N_R$,
with a spin accumulation only in the left lead, $\delta \mu_{\rm L} \equiv \delta \mu  \ne 0$, $\delta \mu_{\rm R} = 0$, and with an applied voltage $eV\equiv \mu_{\rm L}-\mu_{\rm R}$.
Current conservation ensures that
${S}_{\rm RR}={S}_{\rm LL}=-{S}_{\rm RL}=-{S}_{\rm LR}$,
and accordingly we only discuss ${S} \equiv {S}_{\rm RR}$ from now on.
Equation \eqref{eq:F} gives
\begin{subequations}\label{eq:Ftemp}
\begin{eqnarray}
{\cal F}^{\sigma \sigma'}_{\rm LR}
&=& 
\left(eV + \sigma \delta\mu \right)
\text{coth}\left[(eV + \sigma\delta\mu)/2k_{\rm B} T\right] \, , \\
{\cal F}^{\sigma \sigma'}_{\rm LL}
&=& \left(\sigma- \sigma'\right) \, \delta\mu \, 
\text{coth}\left[(\sigma- \sigma') \delta\mu/2k_{\rm B} T \right] \, , \qquad \\
{\cal F}^{\sigma \sigma'}_{\rm RL}
&=& {\cal F}^{\sigma' \sigma}_{\rm LR} \, , 
\;\;\;\;\;\;\;\;\; {\cal F}^{\sigma \sigma'}_{\rm RR} = 2 k_{\rm B}T \, ,
\end{eqnarray}
\end{subequations}
while Eq.~\eqref{eq:chaotic} gives
\begin{equation}\label{TRR}
\langle \Tc{\rm RR}{\gamma \delta}{ab}\rangle_{\rm chaotic} 
= (N/4) \delta_{ab} \left( \delta_{a0}+2\delta_{\gamma\delta} \right).
\end{equation}
In the limit of zero temperature, we get the ensemble averaged zero-frequency noise power as
\begin{equation}
\begin{split}
\langle {S} \rangle = & (1/4) N \left( |eV+\delta\mu|+|eV-\delta\mu|+|\delta\mu| \right) \; .
\end{split}
\label{eq:zeroT}
\end{equation}
This function is plotted in Fig.~2(a). The spin accumulation manifests itself as a change in the slope of the noise at a crossover voltage $|eV|=|\delta\mu|$, with a saturation at $\langle S \rangle = 3/4 \times N |\delta \mu|$
for $|eV| < |\delta \mu|$, turning into $\langle S \rangle = N/4 \times (2 |eV| + |\delta \mu|)$ for $
|eV| > |\delta \mu|$. For $\delta \mu = 0$, we reproduce the result 
$\langle S \rangle = 2 e F I$ with $F=1/4$, valid  
for chaotic ballistic systems~\cite{bla00}. The abrupt change in slope at $|eV|=|\delta \mu|$ is smoothed out at finite temperature. This is shown in Fig.~2(b), where we plot the finite temperature analytic formula for $\langle S \rangle$ obtained from Eqs.~(\ref{eq:mother}),
(\ref{eq:Ftemp}), and (\ref{TRR}). The crossover from low bias, $|eV| < |\delta \mu|$, to high bias, 
$|eV| > |\delta \mu|$, is still extractable from $\partial^2 S/\partial V^2$, as illustrated in 
Fig.~2(c). The second derivative reaches its maximum close to $|eV| = |\delta \mu|$ as long as
$k_{\rm B}T \lesssim |\delta \mu|$. 

\begin{figure}
\includegraphics[width=8cm]{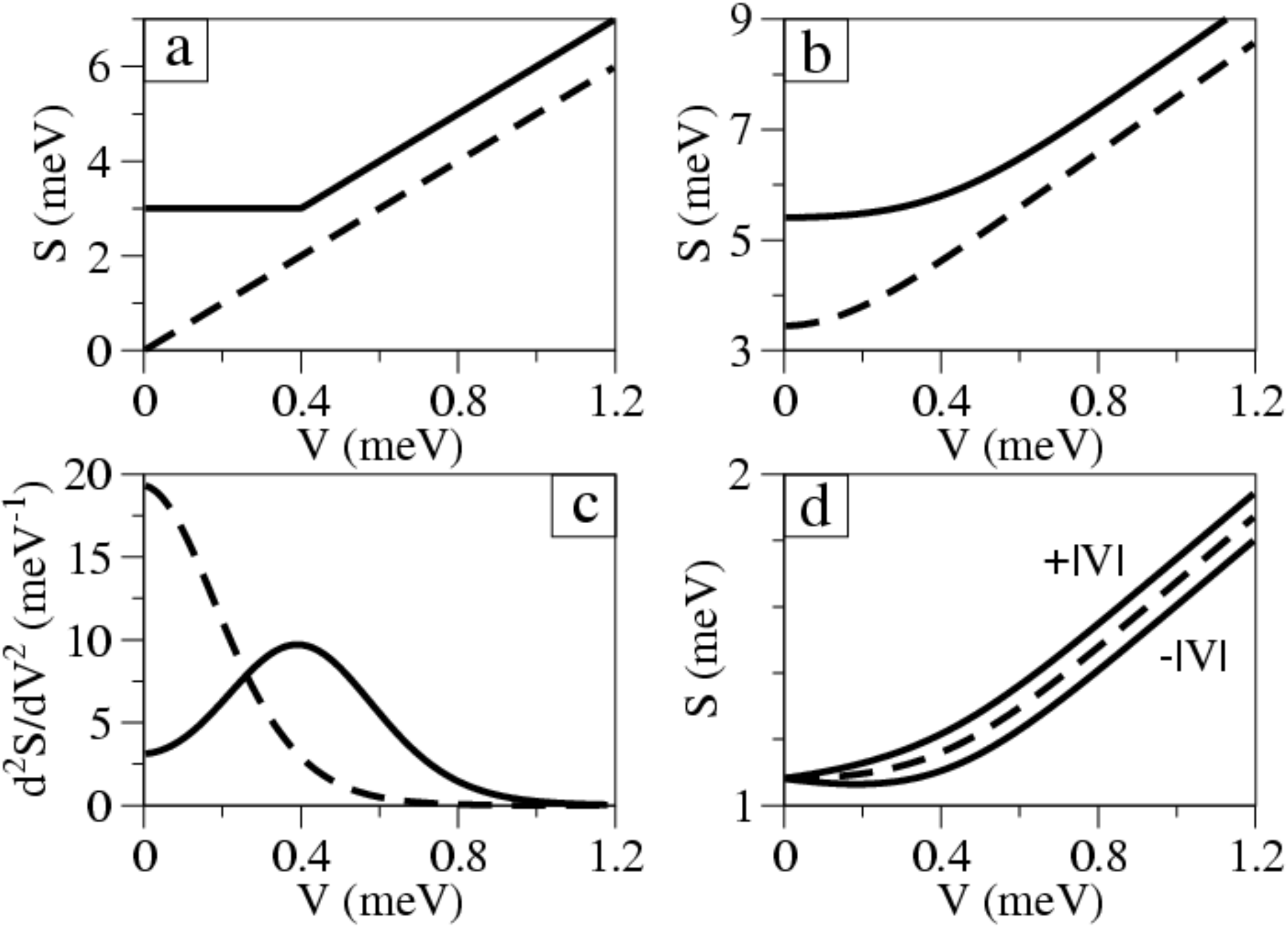}
\caption{\label{fig2} Current noise in a two terminal conductor vs.~applied bias voltage for a spin accumulation of $\delta\mu = 400$ $\mu$eV in a single lead (solid lines) or no spin accumulation in either lead (dashed lines). (a) $T=0$ K and $N=10$. (b) $T=1$ K and $N=10$. (c) Second derivative of the data in panel (b). (d) Typical asymmetry in the current noise as a function of applied bias for $T=1$ K and $N=2$.}
\end{figure}

For zero applied voltage, $V=0$, we get
\begin{equation}
\begin{split}
\langle {S} \rangle = & (11/4) N k_{\rm B}T + (1/4) N \delta\mu \\
&\times \left[ 2\coth(\delta\mu/2k_{\rm B}T)+\coth(\delta\mu/k_{\rm B}T) \right] \, .
\end{split}
\label{eq:zeroV}
\end{equation}
In the low temperature limit, $k_{\rm B}T \ll |\delta\mu|$, the noise due to the spin accumulation decouples from the thermal noise, allowing for the measurement of $\delta \mu$ by varying the temperature. In the opposite limit, $k_{\rm B}T \gg |\delta\mu|$, we recover the standard result for the Johnson-Nyquist noise, $S=4 k_{\rm B}T G$.

So far we have shown how a spin accumulation can be quantitatively extracted from the ensemble averaged
current noise. According to Eq.~(\ref{eq:chaotic}), the average $\langle \Tc{\rm RR}{\rm LR}{z0} \rangle$ vanishes. However, 
individual samples might exhibit a nonzero $\Tc{\rm RR}{\rm LR}{z0}$, which, quite importantly, generates a contribution to the noise that is antisymmetric in the bias voltage. Using Eq.~(\ref{eq:mother}) we get, at zero temperature,
\begin{equation}
\delta S \equiv
{S}(V)-{S}(-V)= 2 \, 
\Tc{\rm RR}{\rm LR}{z0}
\big\{ |eV+\delta\mu|-|eV-\delta\mu| \big\} \, ,
\label{eq:antisymmetric}
\end{equation}
  while at high temperature the effect is washed out, as expected: $\delta S(V) = 4/3 \times T^{z0}_{\rm LRRR} \times \delta \mu \,{\rm e}V / k_{\rm B} T$.  

We estimate the magnitude of this asymmetry in a typical mesoscopic sample by calculating the 
root mean square of
$\Tc{\alpha \beta}{\gamma \delta}{z0}$.
Again, using the method of Ref.~\cite{Bro96}, we find that in chaotic ballistic systems
\begin{equation}
\langle {\rm var}\, 
\Tc{\rm RR}{\rm LR}{z0} 
\rangle_{\rm chaotic} = 1/128 + {\cal O}(N_{\rm T}^{-1}) \, .
\label{eq:rms}
\end{equation}
Accordingly, one has a typical asymmetry of
$\delta S_{\rm typ} = {\rm rms}(\delta S) = |eV|/2 \sqrt{2}$ at low voltages and $\delta S_{\rm typ} = |\delta \mu|/2 \sqrt{2}$ at higher voltages. This typical noise asymmetry is illustrated in Fig.~2(d). 
Interestingly, the asymmetry renders the noise smaller at finite voltage than at $V=0$.
A noise asymmetry was reported in Ref.~\cite{heidi} in systems with broken time-reversal
symmetry in the nonlinear regime. The mechanism for this asymmetry is, however,
different here. 

Because the asymmetry does not scale with the number of channels, while the total noise does, 
we predict that it is more evident in systems with few channels. The next order contributions tend to somewhat reduce the leading order result in Eq.~\eqref{eq:rms}. This is most pronounced at $N=1$, where time-reversal symmetry requires 
that $\Tc{\rm RR}{\rm LR}{z0}$ vanish identically. This is analogous to 
the vanishing of $\text{Tr}\left[
\left(\mathbb{1}_\beta\otimes\sigma_\beta^a\right)
s_{\alpha\beta}^\dagger s_{\alpha\beta}\right]$ found in Ref.~\cite{kis}.
Our calculations therefore suggest that the asymmetry is best visible for $N=2$.

The method of Ref.~\cite{Bro96} can also be applied to diffusive systems with an 
elastic mean free path much smaller than the linear system size, $\ell \ll L$. 
One obtains
\begin{eqnarray}
\langle 
\Tc{\rm RR}{\rm LL}{ab}
\rangle_{\rm diffusive} &=& \delta_{ab} \delta_{a0} (4/3) N \ell/L \, ,\\
\langle 
\Tc{\rm RR}{\rm LR}{00}
\rangle_{\rm diffusive} &=& (2/3) N \ell/L \, ,\\
\langle {\rm var} \,
\Tc{\rm RR}{\rm LR}{z0}
\rangle_{\rm diffusive} &=& (2/35) \ell/L \, .
\end{eqnarray}
This gives, in particular, for $T = 0$
\begin{equation}
\langle {S} \rangle = (2N \ell /3 L) \left( |eV+\delta\mu|+|eV-\delta\mu|+2|\delta\mu| \right) \; ,
\label{eq:zeroTdiff}
\end{equation}
and for $V=0$ 
\begin{equation}
\begin{split}
\langle {S} \rangle = & (4 N \ell / 3 L) \{ 3k_{\rm B}T + \delta\mu  
\\ & \times 
\left[ \coth(\delta\mu/2k_{\rm B}T)+\coth(\delta\mu/k_{\rm B}T) \right] \} \, .
\end{split}
\label{eq:zeroVdiff}
\end{equation}
Comparing Eqs.~\eqref{eq:zeroT} and \eqref{eq:zeroV} with Eqs.~\eqref{eq:zeroTdiff} and \eqref{eq:zeroVdiff} we see that after the substitution $N\to N \ell/L$, the noise averages for chaotic and diffusive conductors differ only by prefactors of order one. 

With the above results, we now evaluate the ratio of a typical noise asymmetry to the ensemble averaged noise. At $|eV|=|\delta\mu|$, where this ratio is maximal, we get, at zero temperature,
\begin{equation}
\delta S_{\rm typ}/\langle S \rangle =
(1/N) \times \left\{
\begin{array}{ll}
\sqrt{2}/3\, , & {\rm  chaotic} \, ,\\
\sqrt{9 L /70 \ell} \, , & {\rm diffusive} \, .
\end{array}
\right. \\
\,
\label{eq:merit}
\end{equation}
Because metallic diffusive wires have $N\gg L/\ell$, we see that a chaotic system is better suited for detection of spin accumulation from the noise asymmetry.

We finally comment on the case of a spin dependent number of channels, $N_{\alpha\uparrow}\neq N_{\alpha\downarrow}$, which occurs for large enough
spin accumulations, $\delta \mu_\alpha/\mu_\alpha > 1/N_\alpha$   and  breaks   time-reversal symmetry. Equation \eqref{eq:chaotic} becomes
\begin{equation}
\begin{split}
\langle 
\Tc{\alpha \beta}{\gamma \delta}{ab} 
\rangle = & 
N_\alpha^0 N_\beta^0 (N_{\rm T}^0)^{-2} \, \left\{ 
N_\gamma^a N_\delta^b \left( \frac{\delta_{\alpha\beta}}{N_\alpha^0}-\frac{1}{N_T^0}\right) \right. \\
& \left.
+ \delta_{\gamma\delta} \left[ N_\gamma^0 \delta_{ab} + N_\gamma^z (1-\delta_{ab}) \right]  \right\} \,,
\end{split}
\label{eq:chaotic2}
\end{equation}
with $N^{0/z}_\gamma=N_{\gamma\uparrow} \pm N_{\gamma\downarrow}$. 
Interestingly, Eq.~\eqref{eq:chaotic2} implies a finite {\it average} asymmetry $\langle \delta S \rangle = {\cal O} (N_\gamma^z)$.


In our derivation of Eq.~\eqref{eq:mother}, we neglected the energy dependence 
of the scattering matrix. This is legitimate as long as the expression in 
brackets in Eq.~(\ref{eq:noise}) is finite only in a narrow energy range. 
When this is not the case, the noise asymmetry will be damped even in individual samples $\delta S_{\rm typ} \rightarrow 0$, unless the spin accumulation is large enough that $N_{\rm L \uparrow} \neq N_{\rm L \downarrow}$. 
Simultaneously, Eq.~\eqref{eq:mother} may still give $\langle S \rangle$ provided one substitutes ${\cal T} \to \langle {\cal T} \rangle$. This is legitimate as long as the response is linear, meaning the applied voltages do not change the electrostatic profile of the conductor, and no substantial energy relaxation takes place in the system. We finally note that, in presence of dephasing, the noise asymmetry determined by Eq.~\eqref{eq:rms} is algebraically damped, in the same way as
conductance fluctuations are.\cite{wht} We thus believe that the noise asymmetry we predict is observable even when dephasing is taken into account.

We thank    Markus B\"uttiker     and Eugene Mishchenko for interesting discussions. This work has been
supported by NSF under Grant DMR-0706319.

\end{document}